# Pros and Cons Gamification and Gaming in Classroom

*Iulian Furdu*
"Vasile Alecsandri" University of Bacău, Bacău, Romania
ifurdu@ub.ro

*Cosmin Tomozei*
"Vasile Alecsandri" University of Bacău, Bacău, Romania
cosmin.tomozei@ub.ro

*Utku Köse*
Uşak University, Uşak, Turkey
utku.kose@usak.edu.tr

**Abstract**
The aim of the current work is to assess the challenges that gamification in education are facing nowadays. Benefits and disadvantages of using gamification in classroom are both discussed to offer a clearer view on the impact of using gamification within learning process. Exploratory study –cases are provided to investigate the relation between motivation and engagement of the students and gamification in training. Following this idea, a survey was conducted to assess how students' behavior and motivation is affected by introducing a single, specific gamification element during a semester learning process. To stimulate competition among students, a ranking type plugin was introduced within the university learning management system used for extramural education. The results prove that motivation decreases by comparison to the previous semester.

## 1. Introduction

Teachers have to use different teaching methods and approaches that allow students to be active participants with strong motivation and engagement to their own learning, and new approaches and techniques in order to implement active learning. Gamification in training is one of these trends.

Introduced in 2002 by Nick Pelling (Pelling, 2011), the gamification did not gain popularity until 2010 when a wide process to incorporate game elements in software was started, marking the point from which big companies like Microsoft, SAP, Deloitte will start using gamification techniques in various applications (Silverman, 2011).

Even if the term gamification suffers various overlaps, it is widely accepted that gamification represents the integration of "game-based mechanics, aesthetics and game thinking to engage people, motivate action, promote learning, and solve problems" (Kapp, 2012), in non-game contexts. We are living in an era in which learners grew up as digital natives and have different learning styles, and new attitude to the learning process. These realities require modern pedagogical paradigms and trends in education (Kiryakova, 2014), in order to fulfill students' requirements, needs, and preferences, to keep them engaged and motivated during their learning process, and gamification is a trend aiming to maintain an active learning. Apart from education (Menezes, C. C. N., & De Bortolli, R., 2016), gamification has found its way into domains like marketing, (Van Grove, J., 2011), knowledge management (Shpakova, A, Dörfler, V. & Macbryde, L., 2016), politics (Rethinking Elections With Gamification, 2015), health (Menezes J.Jr., Gusmão, C., &Machiavelli, J., 2013), road safety (Blohm, I. & Leimeister, J. M., 2013), etc., and analysts predict it will become a 11.10 billion dollar industry by 2020, (Gamification Market, 2016).

This paper is organized as follows: section 2 provides an overview on gamification basic concepts and trends by means of different study cases and examples, section 3 describes some pros and cons of using gamification, next section describes a survey taken to assess how students' behavior and motivation is affected by introducing a new element of gamification (among others already introduced) within the learning management system platform used in the teaching process for bachelor students in computer science. The last section concludes and summarizes the work.

## 2. Gaming and gamification: an overview and basic concepts

Both, games and gamification are rewarding for the educational system and for the learning experience in general. The main differences between game-based learning and gamification imply, firstly, that using gamification does not involve adapting the content to fit the game story and rules as in game-based learning. On the other hand, gamification is used to transform the learning experience into an educational game by using game elements to motivate and keep the students active (usually by a system of rewards or by indicating their level of performance), while in game-based learning activities, games are used to achieve skills or knowledge. According to (Lee, J. J. & Hammer, J., 2011) "understanding the role of gamification in education, means understanding *under what circumstances* game elements can drive learning behavior". This understanding can be derived from various gamification projects. Instead of receiving traditional grades, students earn "experience points" for completing assignments (Laster, J, 2010). During a semester, this lead to a class average of B instead of a previous C. In New York, in a charter school, game designers have participated together with teachers to provide a new curriculum, for providing students with playful activities during the whole day (Corbett, 2010). Improving students` activity during lectures can be obtain by gamification techniques, but further research needs to be done to more accurately match gamification types to learning styles (Rapeepisarn et al. 2008).

Presenting gamification mechanics during classes by implementing them into grade system can be easily obtained by using eLearning environments hybridized with immersive interactive scenarios, like in Lifesaver- a learn by doing model to teach the basic steps in responding to a situation where a person suffers a heart attack or choking (Gamification in eLearning, 2017). The proper use of narrative layers can improve engagement of user and points can be gained using short assignments (missions). The students can choose the assignments as they like to obtain enough points to pass the classes. Obviously, for harder tasks they will get more points, but none of the tasks are obligatory.

Other gamification elements include avatars, badges, levels, reputation level, tasks, etc. Details are presented in Figure 1.

| Game-design elements | | Motives |
|---|---|---|
| Game mechanics | Game dynamics | |
| Documentation of behavior | Exploration | Intellectual curiosity |
| Scoring systems, badges, trophies | Collection | Achievement |
| Rankings | Competition | Social recognition |
| Ranks, levels, reputation points | Acquisition of status | |
| Group tasks | Collaboration | Social exchange |
| Time pressure, tasks, quests | Challenge | Cognitive stimulation |
| Avatars, virtual worlds, virtual trade | Development/organization | Self-determination |

*Figure 1. Game-design elements and motives (Blohm, I. & Leimeister, J. M., 2013).*

Making the rewards for accomplishing tasks visible to other players or providing leaderboards are ways of encouraging players to compete.

## 3. Pro and cons gamification

An effective gamification concept is one that captures and retains learners' attention, engages, entertains and challenges them, and finally teaches them. According to Andrew Phelps gamification is "in its really early days, and we're still drilling into how and why this stuff works and what makes it effective." (Pros and Cons Gamification in Classroom, 2013). Despite this uncertainty, some benefits of using gamification in the instructional process are (Gamification In eLearning, 2017):

A *better learning experience* is obtained by combining "fun" with learning during the game.

A good gamification strategy will make participants more active and high levels of engagement will increase feedback and retention.

*Instant feedback*. Since gamification *provides metrics* it can be easily seen, as trainer, how a participant is progressing. From the students' perspective, tests and assignments, as well as all other activities provide different levels/ways of feedback, so that learners know what they know or what they should know.

*Better learning environment.* The learning experience is personalized; the learners could evolve in their own rhythm, in a safe way. Gratification system provides an effective, informal learning environment that helps learners practice real life situations and challenges.

Gamification is about a lot more than just surface level benefits granted by points, badges, reputation level as it can *catalyze behavioral change*, especially if combined with the scientific principles of cyclical learning and ensuring retention.

Gamification is versatile since, by using it, *most learning needs can be fulfilled*, including product sales, customer support, soft skills, awareness creation, etc., resulting a *performance gain for organizations.*

By the other hand some *drawbacks of using gamification* in an excessively or wrong way must be considered. By *making play mandatory*, gamification might create rule-based experiences that feel just like school. The effort, not mastery, should be rewarded, and the students should learn to *see failure as an opportunity*, instead of becoming unmotivated or fearful. Activities need to be designed so that students can repeat them in case of an unsuccessful attempt (Kiryakova, G., Angelova, N. & Yordanova, L., 2014). Feedback can be used as a correction of students' actions and *should be a stimulus* to their further activities. Also, the trainers should balance metrics with real engagement.

The design of the challenges and the setting of the content have to be carefully considered in order to make it as neutral as possible while not seeming trivial and boring.

According to Kathy Sierra, a popular technology blogger, author and game developer, rewards "should be left at the classroom door" (Gamification in the Classroom, 2014) "a well-designed game only deploys certain mechanics to *support* an intrinsically rewarding experience". If the experience is removed but the mechanics kept, the users psychology changes so that, in essence, it "uses mechanics to drive mechanical behaviors" with little or no gain for the educational process. Nevertheless, motivators like points, badges, leaderboards are not effective for students who aren't naturally competitive, and if these elements will have a central role, the students will finally lose their interest.

### 4. A survey on introducing Moodle ranking block within extramural education

A study was conducted aiming to assess the impact of introduction of ranking block plugin as a gamification element within Moodle learning management system (Ranking block Moodle, 2017). We mention that the Moodle platform, version 3.2 is dedicated to extramural and distance learning. It supports various gamification elements such as avatars, badges, leaderboard, levels, displaying quiz results or progress bars. The ranking block plugin was introduced and configured to be available for procedural programming course activities, at the beginning of the first semester of 2016, which starts in October. It displays a course leaderboard visible to all users as a way of obtaining recognition from other users. It is based on points instead of badges and it can monitor included activities based on accumulated points. The experiment involved first year bachelor students in computer science (32 students, extramural education) from UVAB University (www.ub.ro) who are using the Moodle platform in their tutorial based activities. The main page of UVAB Moodle platform is presented in Figure 2.

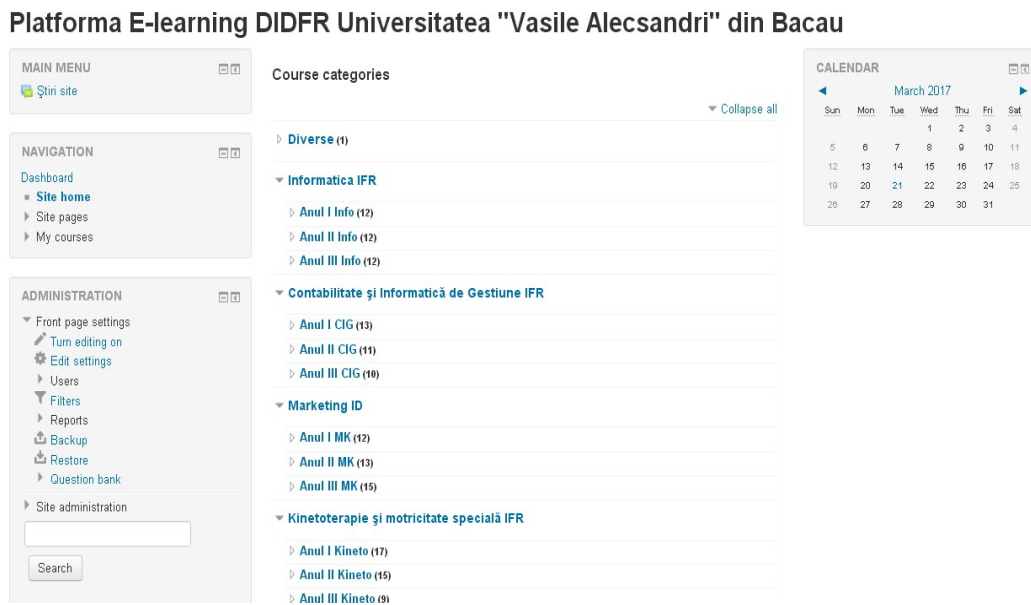

*Figure 2. Course categories within UVAB University Moodle platform, (portal-eifr.ub.ro)*

In the previous semester completion tracking was enabled and set to show activities as complete when conditions are met. Also quiz and lesson completions were set to require a passing grade. Badges and progress bar plugin were also available.

At the end of the second semester a questionnaire-based survey was conducted to evaluate the impact of introducing the Moodle ranking plugin as a gamification element. The most relevant question for this purpose was "Does the ranking board motivated you in your studies?".

Consequently, the results obtained for this item are to be presented. The level of response is 87,5%; the answers to the above question are generally positive, as the results from Table 1 proves, but, despite this fact, the number of platform accesses decreases between the two semesters.

Table 1. Motivational status

| # | "Does the ranking board motivated you in your studies?" | f | % |
|---|---|---|---|
| 1 | Yes | 12 | 42.8 |
| 2 | Partially | 6 | 21.4 |
| 3 | No | 10 | 35.7 |
| | Total respondents | 28 | 100 (87,5%) |

Figure 3 shows the number of course/activities accesses, per week, for the first and the second semester. Combining this with a lower rate of activities completion (not shown here) for the second semester, it is empirically proved that the introduction of the ranking block did not have the assumed motivational effect for students. A good explanation could be the fact that the top 4 students were at the end of semester far ahead from the others, which, most probably weakened the group motivation.

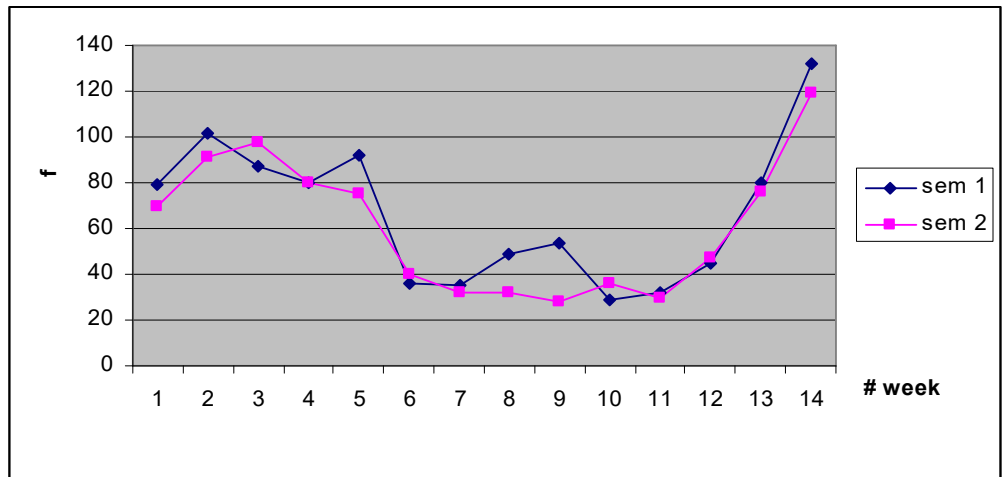

*Figure 3. Number of platform visits/week during first and second semester*

Graphic trends are similar in each 14 weeks semester. In the first month, students are receiving their tasks and usually they access the platform more than in the rest of semester, except the last weeks when they start to prepare for exams and to fulfill their end of semester tasks. The students' workload was similar in each semester, so that this aspect could not bias the results. However, the adaptation process of first-year students to the university requirements/working style could influence the outcome, but not to a relevant extent. Overall, the decrease in using the platform was of 9,16%, from 932 visits in the first semester to 854 in the second one.

## 5. Conclusions and future work

Benefits and disadvantages of using gamification in classroom were discussed in order to offer a clearer view on the impact of using gamification within learning process. The relation between motivation and engagement of the students and gamification in training was explored by study-cases. A survey was conducted to assess how students' behavior and motivation is affected by introducing for a semester, a ranking type plugin within the university learning management system used for extramural education. The results empirically proved that motivation decreases by introducing leaderboards in given circumstances (students, year, and specialization). To increase the consistency of the study, future work will assess if the negative trend recovers after eliminating the ranking plugin, and explore other gamification techniques to improve learning experiences for students.

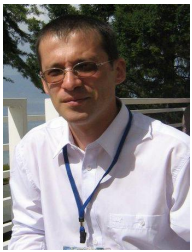

**Iulian Furdu** received his BSc in Mathematics and Physics (1995), PhD in Computer Science (2010) from "Gheorghe Asachi" Technical University of Iași, Romania. His current research interests include different aspects of Artificial Intelligence, GIS, e-Education. He contributed to 6 research and development projects, (co-)authored 10 books and book chapters, and more than 40 research papers. ResearcherID: C-6818-2015.

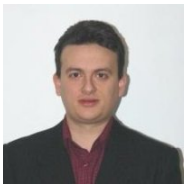

**Cosmin Tomozei**, PhD in Cybernetics, is Lecturer in Computer Science, at Vasile Alecsandri University of Bacau, Romania. His main research areas are: object oriented programming, functional programming in Lisp and F#, software reengineering and distributed applications development.

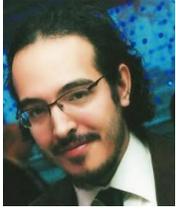

**Utku Köse** received the B.S. degree in 2008 from computer education of Gazi University, Turkey as a faculty valedictorian. He received M.S. degree in 2010 from Afyon Kocatepe University, Turkey in the field of computer and D.S. / Ph. D. degree in 2017 from Selcuk University, Turkey in the field of computer engineering. Between 2009 and 2011, he has worked as a Research Assistant in Afyon Kocatepe University. Following, he has also worked as a Lecturer and Vocational School - Vice Director in Afyon Kocatepe University between 2011 and 2012. Currently, he is a Lecturer in Usak University, Turkey and also the Director of the Computer Sciences Application and Research Center at the same university. His research interest includes artificial intelligence, optimization, chaos theory, distance education, e-learning, computer education, and computer science.